\documentclass[10 pt,journal]{IEEEtran}
\usepackage{amsmath,amsfonts}
\usepackage[linesnumbered,ruled,vlined]{algorithm2e}
\usepackage{array}
\usepackage[caption=false,font=footnotesize,labelfont=sf,textfont=sf]{subfig}
\usepackage{caption}
\usepackage{textcomp}
\usepackage{stfloats}
\usepackage{url}
\usepackage{verbatim}
\usepackage{graphicx}
\usepackage{cite}
\usepackage{multirow}
\usepackage{multicol}
\usepackage[dvipsnames,table]{xcolor}
\usepackage{subcaption}
\usepackage{booktabs}

\usepackage{verbatim}

\newcommand{\ie}{\textit{i.e.}}
\newcommand{\etal}{\textit{et~al.}}
\newcommand{\vect}[1]{\mathbf{#1}}
\newcommand{\norm}[1]{\left\lVert #1 \right\rVert}

\begin{document}

\title{\fontsize{21}{25}\selectfont \textit{CLIF}: Cross-layer LEO-ISL Fingerprinting for Physical and Network Attack Detection in Dense LEO Constellations}

\author{Varun Kohli,~\IEEEmembership{Graduate Student Member,~IEEE,}, Arijit Bhattacharjee,~\IEEEmembership{Member,~IEEE,}, Samar Shailendra,~\IEEEmembership{Senior Member,~IEEE,}, Biplab Sikdar,~\IEEEmembership{Fellow,~IEEE,}
\thanks{V. Kohli is with the Institute for Infocomm Research ($I^2R$), Agency for Science, Technology and Research (A*STAR), 1 Fusionopolis Way, \#21-01 Connexis, Singapore 138632, and the Department of Electrical and Computer Engineering, National University of Singapore, Singapore 117417 (email: kohliv@a-star.edu.sg, varun.kohli@u.nus.edu).}
\thanks{A. Bhattacharjee and B. Sikdar are with the Department of Electrical and Computer Engineering, National University of Singapore, Singapore 117417. (e-mail: b.arijit@nus.edu.sg, bsikdar@nus.edu.sg)}
\thanks{S. Shailendra is the CTO of EdGenAI Technologies, Melbourne, Australia. (e-mail:  samar@edgenai.com.au)}}



\maketitle

\begin{abstract}
Low-Earth Orbit (LEO) mega-constellations such as Starlink by SpaceX and Kuiper by Amazon rely on optical Inter-Satellite Links (ISLs) for autonomous mesh routing to provide low-latency telecommunication, Internet of Things (IoT), and security services globally. As commercial operators and governments deploy increasingly dense constellations and form multi-operator peering coalitions, ISL integrity becomes critical to both commercial availability and national security. However, there is a lack of real-world data for LEO constellations and existing real-time security approaches focus strictly on physical layer security, leaving blind spots in the coverage of network-layer and composite attacks. In this paper, we present a cross-layer, lightweight behavioral fingerprinting framework that fuses onboard physical-layer measurements with network-layer data to detect anomalies at low computational overhead. We construct an orbital simulation covering the first shells of Starlink (1,584 satellites), Kuiper (1,156 satellites), and a joint multi-operator peering scenario (2,740 satellites), injecting ten attack types that span spoofing, traffic manipulation, and routing subversion at varying severity. We evaluate three unsupervised, per-satellite detectors among which our Mahalanobis-distance-based detector achieves 99.5\% recall on Starlink, 99.4\% on Kuiper, and 94.8\% on the multi-operator constellation, while maintaining False Positive Rates (FPR) below 0.7\%. Our results demonstrate that cross-layer feature fusion is not only necessary for comprehensive security of LEO constellations but highly cost-effective for large-scale networks while fitting into the strict onboard energy budgets of resource-constrained satellites.
\end{abstract}


\section{Introduction}
\IEEEPARstart{T}{he} rapid deployment of Low-Earth Orbit (LEO) constellations is reshaping telecommunication, Internet of Things (IoT), and national security across the world. SpaceX's Starlink constellation already operates over 10,000 LEO satellites, with regulatory filings describing a first shell of 1,584~satellites at a 550~km altitude \cite{starlink_fcc}, while Amazon's Project Kuiper constellation consists of 1,156~satellites at 630~km in its first shell \cite{kuiper_fcc}. These constellations rely on optical Inter-Satellite Links (ISLs) operating at C-band (193.4~THz, $\sim$1550~nm) to form an autonomous mesh network in orbit, enabling global coverage of high-throughput services through efficient packet routing at low latency without continuous ground-station contact \cite{handley2018delay, bhattacherjee2018gearing}. To augment national security and achieve sovereignty in space-based telecommunication, governments across the globe are also developing national-level LEO constellations and cross-border coalitions with their allies. Some examples include the EU IRIS2 \cite{gur2025eu}, and OneWeb \cite{kozhaya2024first}.  

The integrity of ISLs is a critical yet largely unaddressed security concern. Unlike the terrestrial infrastructure, ISLs traverse an uncontrolled environment where an adversary can exploit firmware vulnerabilities to compromise satellite nodes \cite{willbold2023space}, intercept and manipulate satellite data in transit \cite{pavur2020tale}, or flood links with Denial-of-Service (DoS) traffic. Liu~\etal~\cite{liu2024dark} recently demonstrated that control-plane signaling storms can block direct-to-cell LEO satellites across entire urban areas. Meanwhile, Yanev~\etal~\cite{yanev2025secure} showed that LEO routing fabrics are fragile under node and link failures, with segment-based rerouting reducing message drops by up to 30\% over neighbor-based methods, making routing-layer attacks particularly impactful. LEO constellations are evolving toward high-density, multi-operator peering. As the attack surface expands across constellation boundaries, per-satellite and ISL-level attack detection becomes a necessity. While the 3GPP Non-terrestrial Network (NTN) standardization effort \cite{3gpp_ntn} addresses ground-to-satellite communication security, it does not address intrusion detection for ISLs.

\subsection{Gaps in the State of the Art}

Securing LEO-NTN requires adapting to rapid topological changes and resource constraints. Existing security approaches for LEO constellations are limited and include Physical-layer Authentication (PLA) \cite{topal2022physical,kohli2026emma} for spoof detection, cryptographic verification \cite{farrea2025zero} to identify unauthorized satellites, and structural network resilience approaches \cite{yuan2026hydra,li2024robustness} for overall network resilience. We identify three principal gaps: \textbf{First,} there is no cross-layer ISL security framework. PLA methods \cite{topal2022physical,kohli2026emma} detect rogue satellites but are blind to network-layer manipulation. Cryptographic approaches \cite{farrea2025zero} reject unauthorized entities but miss compromised insiders with valid credentials. Structural resilience approaches \cite{yuan2026hydra,li2024robustness} identify which nodes to harden but operate solely on network topology, overlooking physical-layer ISL measurements. Meanwhile, Liu~\etal~\cite{liu2024dark} and Yanev~\etal~\cite{yanev2025secure} have shown that LEO constellations face real signaling and routing-layer threats. So far, no detection mechanism exists that spans both physical and network layers to catch composite attacks. \textbf{Second,} Prior evaluations focus on single attack types \cite{topal2022physical,farrea2025zero} but do not share their data, use simplified topologies with up to 60 satellites \cite{kohli2026emma}, or study structural robustness without per-link ground truth \cite{yuan2026hydra,li2024robustness}. There is no publicly available dataset capturing realistic ISL under attack. \textbf{Third,} there is no existing work that covers multi-operator ISL threats such as inter-constellation traffic hijacking.

\subsection{Contributions}

This paper makes the following contributions:

\begin{enumerate}
  \item \textbf{Cross-layer behavioral fingerprinting framework.} We propose an unsupervised, per-satellite anomaly detection framework for monitoring ISL connections using three physical-layer features (range mismatch, velocity mismatch, and direction mismatch) with nine network-layer features (throughput, utilization, sender and receiver queue depth, queue delay, total latency, packet loss rate, sender neighbor count, and traffic asymmetry) into a 12-dimensional cross-layer feature space. Each satellite hosts one lightweight, unsupervised model for real-time detection, requiring no inter-satellite coordination. 
  \item \textbf{ISL simulation and dataset with comprehensive threat modeling.} Based on public information on the first shells of Starlink and Kuiper, we build a physics-grounded simulation that propagates orbits via SGP4, constructs dynamic plus-grid ISL topologies, and overlays congestion-aware Dijkstra routing with Poisson traffic and M/M/1/K queuing. We implement ten attack types spanning physical-layer deception, network-layer manipulation, cross-layer composites, and multi-operator exploitation, each at varying severity, capturing attacks of different types and difficulty. Our dataset consists of 24 hours of simulation data for Starlink (1,584 satellites), Kuiper (1,156 satellites), and a joint Starlink-Kuiper multi-operator network (2,740 satellites) and is published on IEEE Dataport \footnote{The dataset is available on https://dx.doi.org/10.21227/p5zv-b264, and the simulation's code will be made available on acceptance.} to drive future research in this domain.
  \item \textbf{Dense constellations and multi-operator evaluation.} We evaluate across the above three constellation deployments, demonstrating that our framework generalizes across configurations and operator boundaries with marginal capability deterioration.
\end{enumerate}

\begin{table*}[t]
\centering
\caption{Related works on physical- and network-layer security of LEO constellations.}
\label{tab:related}
\renewcommand{\arraystretch}{1.2}
\resizebox{\linewidth}{!}{
\begin{tabular}{|c|c|l|c|c|c|c|c|}
\hline
\textbf{Method} & \textbf{Layer} & \textbf{Attack Types} & \textbf{Granularity} & \textbf{Mode} & \textbf{Constellation (\# Sat)} & \textbf{Multi-operator} & \textbf{Dataset}\\ \hline
\cite{topal2022physical}                & Physical                            & Rogue, Spoofing                                                                                                                                                                          & ISL                                       & Real-time     & N/A (6-10) & No & No                     \\ \hline
\cite{kohli2026emma}                & Physical                            & Rogue, Spoofing                                                                                                                                                                          & ISL                                       & Real-time         & Iridium (60) & No & Yes                 \\ \hline
\cite{farrea2025zero}                & Physical                       & Authentication, Sybil                                                                                                                                                                    & Satellite                                 & Real-time        & N/A (2) & No & No                  \\ \hline
\cite{li2024robustness}                & Network                             & Cascading failure prediction                                                                                                                                                                       & Satellite                                & Offline        & N/A (100) & No & No                    \\ \hline
\cite{yuan2026hydra}                & Network                             & Cascading failure prediction                                                                                                                                                                        & Satellite                                 & Offline           & Starlink (9,300) & No & No                 \\ \hline
\cite{yanev2025secure}                & Network                             & Link/node failure prediction                                                                                                                                                                        & Path                                      & Real-time       & \begin{tabular}[c]{@{}c@{}}Iridium (66)\\ Starlink (1,584)\end{tabular} & Yes & No                     \\ \hline
\textbf{Ours}                & \begin{tabular}[c]{@{}c@{}}Physical\\+\\Network\end{tabular}   & \begin{tabular}[c]{@{}l@{}}\textbf{10 attack types}\\ Rogue/spoofing, Blackhole,\\ Sinkhole, Wormhole, Sybil \\ Rogue Sybil, Rogue Sinkhole, \\ DoS, Inter-constellation hijack\end{tabular} & Satellite                                 & Real-time     & \begin{tabular}[c]{@{}c@{}} Kuiper (1,156) \\ Starlink (1,584) \\ Multi-operator (2,740)\end{tabular} & Yes & Yes                     \\ \hline
\end{tabular}}
\end{table*}

\section{Related Work}
\label{sec:related_work}

This section presents related works in NTN and LEO constellation security, highlighting gaps that motivated the work in this paper. Table~\ref{tab:related} summarizes the coverage of prior approaches across five dimensions: observation layer, attack coverage, detection granularity, constellation scale, and operational mode.

\subsection{Satellite Communication Security}

There is considerable work that secures satellite-to-ground communication. The 3GPP NTN framework \cite{3gpp_ntn} extends 5G to satellite access, addressing authentication and encryption on the ground-to-satellite radio interface, but treats the satellite payload as a transparent or regenerative relay and does not specify security mechanisms for the inter-satellite segment. Salim \etal \cite{salim2024cybersecurity} provided a comprehensive survey of cyber-attacks across space, ground, and links segments of NTNs, noting that ISL-layer security remains largely unaddressed. Meng \etal \cite{meng2025comprehensive} surveyed satellite authentication methods across five categories, namely, cryptographic, blockchain-based, orbital parameter-based, protocol-based, and physical hardware-based, and found no existing method that address cross-layer ISL anomaly detection. Earlier surveys \cite{manulis2021cyber} discuss threats including eavesdropping, jamming, and spoofing on feeder and user links, but note ISL-layer threats only as an open problem. Among research focusing on attacks, Liu \etal \cite{liu2024dark} demonstrated a control-plane attack in which a terrestrial adversary triggers signaling storms that block direct-to-cell LEO satellites across entire urban areas, while Willbold \etal \cite{willbold2023space} reverse-engineered operational satellite firmware and identified exploitable vulnerabilities, confirming that attacks on satellite systems are practical. \textbf{The common aspect in the above is that satellite communication security does not extend to the ISL-enabled autonomous mesh topologies that characterize modern LEO constellations.}

\subsection{LEO Constellation Networking}

The networking properties of LEO mega-constellations have attracted significant research attention. Handley \cite{handley2018delay} demonstrated that Starlink's ISL mesh can achieve lower end-to-end latency than terrestrial fiber for intercontinental paths, motivating the analysis of ISL routing. Bhattacherjee \etal \cite{bhattacherjee2018gearing} compared planned mega-constellation designs and evaluated their network performance against terrestrial infrastructure, showing that ISL topology significantly affects routing performance. Del Portillo \etal \cite{del2019technical} provided a technical comparison of Starlink, Telesat, and OneWeb, establishing the constellation parameters used in subsequent simulation work. Yanev \etal \cite{yanev2025secure} studied rerouting resilience under link and node failures in Starlink and Iridium, finding that segment-based rerouting reduces message drops by up to 30\% compared to neighbor-based methods. These studies establish the networking and resilience context for LEO ISLs but do not address adversarial attack detection on individual links.

\subsection{Physical-Layer Security for ISLs}

PLA exploits signal characteristics that are difficult for an attacker to replicate. In the satellite domain, PLA has primarily targeted ground-to-satellite links. Oligeri \etal~\cite{oligeri2020gnss} exploited Iridium signal fingerprints for Global Navigation Satellite System (GNSS) spoofing detection, demonstrating that satellite transmitter-specific features can be used as fingerprints. For ISLs specifically, Topal \etal~\cite{topal2022physical} proposed a two-phase Doppler-based PLA scheme in which multiple receiving satellites independently compare measured normal Power Spectral Density Samples (NPSDS) against reference values computed from the transmitter's known orbital parameters, then fuse their binary decisions via a majority rule to authenticate the transmitter. While effective for rogue satellite detection with high analytical detection probability, this approach relies solely on Doppler shift, requires multi-satellite coordination, and is by design blind to network-layer attacks. Kohli \etal~\cite{kohli2026emma} proposed EMMA, an ephemeris-assisted multi-task learning model that predicts range, elevation, and azimuth from delay, Doppler and sector measurements, detecting rogues via reconstruction error. EMMA improves on single-feature methods by jointly predicting multiple geometric quantities, but its per-ISL-pair training and direction-dependent geometry limit coverage to stable in-plane links with sufficient training observations.

\subsection{Network Security and Structural Resilience}

Network intrusion detection for satellite networks has largely adapted terrestrial methods. Machine Learning (ML)-based anomaly detection has been applied to satellite communication for fault detection \cite{hundman2018detecting}, however, targeting spacecraft health monitoring rather than adversarial attacks on the data plane. For routing security, Blackhole and Sybil attacks are well studied in Mobile Ad-hoc Networks (MANETs), but detection approaches for terrestrial networks generally assume random, unpredictable node mobility and cannot exploit the deterministic link geometry of orbital constellations, where satellite positions and link durations are known in advance from predictable orbital paths and drift. Cryptographic approaches such as zero-trust authentication \cite{farrea2025zero} provide strong guarantees against unauthorized entities but cannot detect a compromised node that possesses legitimate keys but also manipulates traffic at the network layer.

A separate line of work studies the structural resilience of LEO constellations. Li \etal \cite{li2024robustness} modeled LEO networks as hyper-networks and analyzed cascading failure robustness, finding that peripheral nodes can inflict greater damage than central ones. Yuan \etal \cite{yuan2026hydra} proposed HYDRA, a hypergraph-based framework that identifies structurally critical nodes in Starlink whose failure triggers disproportionate cascading damage, reporting a 24.1\% improvement in identifying high-impact nodes over degree-based metrics. These works identify risky nodes for preventive measures but do not model specific attack types and do not provide per-ISL real-time detection of ongoing attacks. Our work is complementary as it provides cross-layer behavioral fingerprinting for real-time detection of active threats on individual ISLs.

\subsection{Summary}

No existing work fuses physical- and network-layer observations for per-link ISL anomaly detection, no publicly available dataset captures cross-layer data for ISLs under attack at constellation scale, and multi-operator ISL threats remain understudied.   

\section{Preliminaries}
\label{sec:preliminaries}
This section introduces the preliminary concepts, including LEO orbital mechanics, ISL observables, and ephemeris models, that the cross-layer detection framework proposed in this paper is built upon. Table~\ref{tab:notation} collects the key notations.

\subsection{LEO Orbital Mechanics}
\label{subsec:orbital_mechanics}

A LEO satellite at altitude $h$ above the Earth's surface orbits at radius $r = R_\oplus + h$, where $R_\oplus = 6,378.135$~km is the equatorial radius. The circular orbital velocity and period are calculated as follows:
\begin{equation}
  v_{\text{orb}} = \sqrt{\frac{\mu}{r}}, \qquad T_{\text{orb}} = 2\pi\sqrt{\frac{r^3}{\mu}},
  \label{eq:orbital_basics}
\end{equation}
where $\mu = 3.986 \times 10^{14}$~m$^3$/s$^2$ is the Earth's gravitational parameter. As an example, $h = 550$~km yields $v_{\text{orb}} \approx 7.59$~km/s and $T_{\text{orb}} \approx 95.7$~min for the first shell of Starlink satellites. Further, a Walker Delta constellation is parameterized by $(N_o, N_s, i)$, where $N_o$ is the number of orbital planes, $N_s$ is the number of satellites per plane, and $i$ is the inclination. Right Ascension of the Ascending Node (RAAN) $\Omega_k = k \cdot 360^\circ / N_o$ for plane $k$ is uniformly spaced, and within each plane, satellites are phased by mean anomaly $M_j = j \cdot 360^\circ / N_s$ for satellite $j$.

\subsection{Inter-Satellite Links}
\label{subsec:isl_model}

Each satellite maintains $L = 4$ laser ISL terminals in a plus-grid configuration \cite{kohli2026emma} such that two \emph{intra-plane} links connect adjacent satellites in the same orbital ring, and two \emph{inter-plane} links connect to the nearest eligible satellites in neighboring planes. Intra-plane links are static, while inter-plane links are dynamic and recomputed as the constellation evolves.

For an ISL between satellites $s_1$ and $s_2$, the terminal on $s_1$ directly measures three physical quantities: (i)~range $d_{\text{meas}}$, via two-way laser ranging from the round-trip time of a laser pulse, (ii)~Doppler shift $f_{d,\text{meas}}$, the carrier frequency offset measured by the optical phase-locked loop, and (iii) Line-of-Sight (LOS) unit vector $\hat{\boldsymbol{\ell}}_{\text{meas}}$, from the gimbal pointing angles. Radial velocity $v_{r,\text{meas}}$ and one-way propagation delay $\tau_{\text{meas}}$ are derived from Doppler and range, respectively.

Given the Earth-Centered Earth-Fixed (ECEF) position vectors $\vect{r}_1, \vect{r}_2$ and velocity vectors $\vect{v}_1, \vect{v}_2$ of any two satellites, the physical model relating state vectors to ISL observables is:
\begin{equation}
    d = \norm{\vect{r}_2 - \vect{r}_1},
    \label{eq:range}   
\end{equation}
\begin{equation}
    \hat{\boldsymbol{\ell}} = \frac{\vect{r}_2 - \vect{r}_1}{d},
    \label{eq:los} 
\end{equation}
\begin{equation}
    v_r = (\vect{v}_2 - \vect{v}_1) \cdot \hat{\boldsymbol{\ell}},
    \label{eq:radvel} 
\end{equation}
\begin{equation}
    f_d = -\frac{v_r}{c} \cdot f_c,
    \label{eq:doppler}  
\end{equation}
\begin{equation}
    \tau = \frac{d}{c},
    \label{eq:delay}  
\end{equation}
where $c = 299,792,458$~m/s is the speed of light and $f_c = 193.4$~THz is the optical carrier frequency.

\subsection{Ephemeris Model}
\label{subsec:ephemeris}

We assume that a secure, Geostationary Earth Orbit (GEO)-based broadcast ephemeris service provides position and velocity claims $(\vect{r}_k^{\text{eph}}, \vect{v}_k^{\text{eph}})$ for an arbitrary satellite $k$ in space. Applying Eq. \eqref{eq:range}-\eqref{eq:delay} to the ephemeris state vectors yields the predicted observables $d_{\text{eph}}$, $v_{r,\text{eph}}$, and $\hat{\boldsymbol{\ell}}_{\text{eph}}$. Under normal conditions, terminal measurements align with ephemeris predictions up to measurement noise. However, under a rogue satellite attack, the actual transmitter occupies a different orbit, causing $d_{\text{meas}}$, $v_{r,\text{meas}}$, and $\hat{\boldsymbol{\ell}}_{\text{meas}}$ to diverge from their ephemeris-predicted counterparts.

\begin{table}[t]
\centering
\caption{Summary of notation.}
\label{tab:notation}
\renewcommand{\arraystretch}{1.2}
\resizebox{\columnwidth}{!}{\begin{tabular}{cl}
\toprule
\textbf{Symbol} & \textbf{Description} \\
\midrule
\multicolumn{2}{l}{\textit{Orbital mechanics}} \\
$\mu$ & Earth gravitational parameter ($3.986\times10^{14}$~m$^3$/s$^2$) \\
$R_\oplus,\; r,\; h$ & Earth equatorial radius, orbital radius, altitude \\
$(N_o, N_s, i)$ & Walker Delta: planes, satellites/plane, inclination \\
\midrule
\multicolumn{2}{l}{\textit{ISL observables}} \\
$L$ & ISL laser terminals per satellite \\
$d$ & Inter-satellite range (m) \\
$v_r$ & Radial (line-of-sight) velocity (m/s) \\
$f_d$ & Doppler shift (Hz) \\
$\tau$ & One-way propagation delay (s) \\
$\hat{\boldsymbol{\ell}}$ & Line-of-sight unit vector \\
\midrule
\multicolumn{2}{l}{\textit{Mismatch features (physical layer)}} \\
$\delta_{\text{range}}$ & Normalized range mismatch, Eq. \eqref{eq:range_mismatch} \\
$\delta_{\text{vel}}$ & Normalized velocity mismatch, Eq.  \eqref{eq:vel_mismatch} \\
$\delta_{\text{dir}}$ & Direction mismatch, Eq. \eqref{eq:dir_mismatch} \\
\midrule
\multicolumn{2}{l}{\textit{Network model}} \\
$w_{ij}$ & Link routing cost \\
$\rho$ & Link utilization ($= \lambda / \mu_s$) \\
$\mu_s$ & M/M/1 service rate (pkt/s) \\
$P_{\text{loss}}$ & Packet loss probability, Eq.  \eqref{eq:packet_loss} \\
$\bar{\lambda}$ & Nominal per-link traffic arrival rate \\
$\lambda_e$ & Instantaneous traffic rate on link $e$, Eq. \eqref{eq:traffic_rate} \\
$\ell_e$ & ECMP load factor \\
$K$ & Queue buffer capacity (packets) \\
$C_e$ & Link capacity \\
$B$ & Packet size (bytes) \\
$g(\phi)$ & Geographic demand weight, Eq. \eqref{eq:geo_weight} \\
\midrule
\multicolumn{2}{l}{\textit{Cross-layer features (Table~\ref{tab:features})}} \\
$\Gamma_e,\; \rho_e$ & Total throughput, link utilization \\
$Q_{s_1},\; Q_{s_2}$ & Queue depths at source, destination \\
$\tau_q,\; \Lambda_e$ & Queuing delay, end-to-end latency \\
$n_e,\; \eta_e$ & Neighbor count, traffic asymmetry \\
\midrule
\multicolumn{2}{l}{\textit{Graph and network model}} \\
$\mathcal{G}_t = (\mathcal{V}, \mathcal{E}_t)$ & ISL topology graph at time $t$ \\
$\mathcal{E}^{\text{intra}},\mathcal{E}_t^{\text{inter}},\mathcal{E}_t^{\text{cross}}$ & Intra-plane, inter-plane, cross-constellation edges \\
$\alpha$ & Congestion weight in routing cost, Eq. \eqref{eq:link_cost} \\
$\beta_{ij}$ & BGP peering penalty (ms) \\
$v_{\text{track}}$ & Laser terminal tracking velocity limit (m/s) \\
\midrule
\multicolumn{2}{l}{\textit{Detection framework}} \\
$\vect{x}_e \in \mathbb{R}^p$ & Cross-layer feature vector ($p{=}12$) \\
$\mathcal{A}(\vect{x})$ & Anomaly scoring function \\
$\theta$ & Detection threshold \\
$D_M(\vect{x})$ & Mahalanobis distance \\
$\boldsymbol{\hat{\mu}},\; \boldsymbol{\Sigma}$ & Feature mean vector, covariance matrix \\
$y_e \in \{0,1\}$ & Binary detection decision for ISL $e$ \\
$\gamma$ & FPR target (0.005) \\
$\mathcal{N}_i^t$ & Active ISL set for satellite $s_i$ at time $t$ \\
$\mathcal{F}_i^t$ & Alert set of flagged anomalous links \\
\bottomrule
\end{tabular}}
\end{table}

\section{Network and Threat Model}
\label{sec:network_threat}

\begin{figure*}[t]
    \centering
    \includegraphics[width=0.9\linewidth]{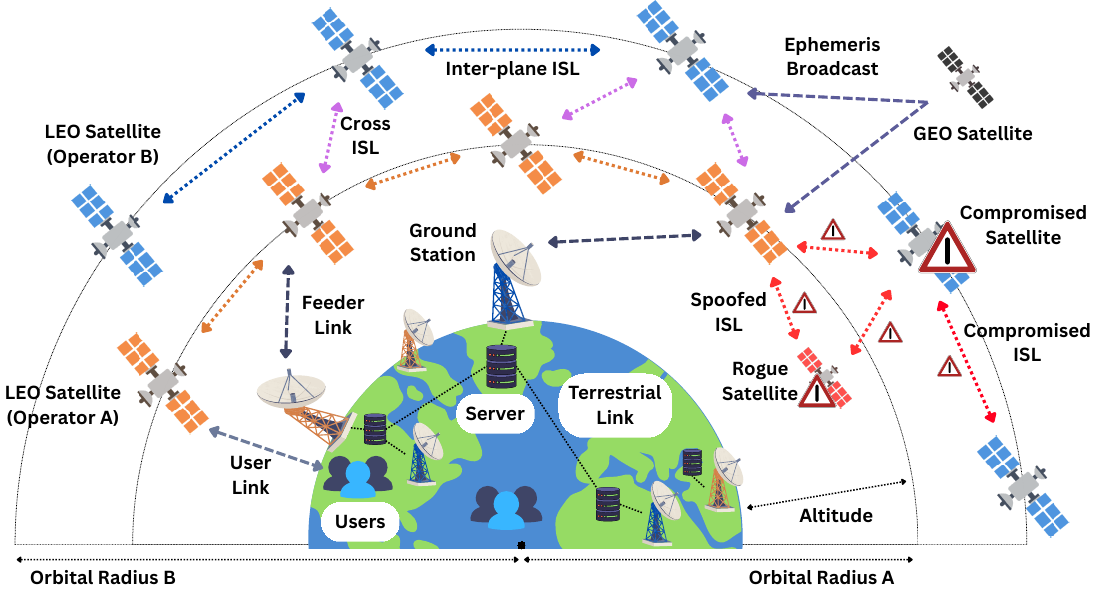}
    \caption{LEO-NTN network with multiple LEO operators.}
    \label{fig:network}
\end{figure*}

This section formalizes the constellation topology as a time-varying graph, specifies the network-layer simulation model, and defines the adversary capabilities and attack taxonomy, focusing primarily on aspects related to LEO constellations. Fig.~\ref{fig:network} depicts a multi-operator LEO-NTN network.

\subsection{Constellation Graph Model}
\label{subsec:graph_model}

We model a LEO constellation as a time-varying undirected graph $\mathcal{G}_t = (\mathcal{V}, \mathcal{E}_t)$ where $\mathcal{V}$ is the set of satellite nodes and $\mathcal{E}_t$ is the set of active ISLs at discrete timestep $t$. In the multi-operator scenario, the vertex set is the union $\mathcal{V} = \mathcal{V}_A \cup \mathcal{V}_B$ of satellites from constellation A and B, respectively. The edge set decomposes as $\mathcal{E}_t = \mathcal{E}^{\text{intra}} \cup \mathcal{E}_t^{\text{inter}} \cup \mathcal{E}_t^{\text{cross}}$ where:

\begin{itemize}
    \item \textit{Intra-plane} $\mathcal{E}^{\text{intra}}$: Each orbital plane forms a static ring. Satellite $j$ in plane $k$ connects to satellites $j-1$ and $j+1(mod\; N_s)$. These links are time-invariant because adjacent co-planar satellites maintain constant relative geometry.
    \item \textit{Inter-plane} $\mathcal{E}_t^{\text{inter}}$: Each satellite connects to the nearest eligible satellite in each of the two adjacent orbital planes. These links are dynamic because the relative positions of satellites in different planes evolve.
    \item \textit{Cross-constellation} $\mathcal{E}_t^{\text{cross}}$: Satellites from different constellations establish peering links when they meet a range criteria. These are also dynamic.
\end{itemize}

\subsection{Network Model}
\label{subsec:network_model}

We now formalize the network-layer behavior on the ISL topology $\mathcal{G}_t$, capturing routing, traffic, and queuing dynamics that characterize realistic constellation operations.

\subsubsection{Routing} 
Following the shortest-path ISL routing established by Handley \cite{handley2018delay} and Bhattacherjee \etal \cite{bhattacherjee2018gearing}, we adopt a congestion-aware link cost that may be recomputed at each routing interval as follows:
\begin{equation}
    w_{ij} = \tau_{ij} + \alpha \cdot \tau_{q,ij} + \beta_{ij},
    \label{eq:link_cost}
\end{equation}
where $\tau_{ij}$ is the propagation delay, $\tau_{q,ij}$ is the queuing delay from the previous interval, $\alpha$ is a congestion weight, and $\beta_{ij}$ is a Border Gateway Protocol (BGP)-like peering penalty for cross-constellation links ($\beta_{ij} > 0$ when operators $o(i) \neq o(j)$, and 0 otherwise), reflecting the operator preference for intra-constellation paths.

\subsubsection{ECMP}
When multiple shortest paths exist, traffic is distributed proportionally according to edge betweenness centrality \cite{feng2022faster}. Each link receives a load factor $\ell_e$ that modulates its share of the aggregate traffic volume.

\subsubsection{Traffic model}
Aggregate ISL traffic arrivals follow a Poisson process, consistent with the well-established convergence of superposed independent streams on heavily multiplexed backbone links \cite{arfeen2019role}. Let $\bar{\lambda}$ denote the nominal per-link arrival rate. The instantaneous rate on link $e$ is defined as follows:
\begin{equation}
    \lambda_e = \bar{\lambda} \cdot \ell_e \cdot g(\phi_e),
    \label{eq:traffic_rate}
\end{equation}
where $\ell_e$ is the Equal-Cost Multi-Path (ECMP) load factor and $g(\phi_e)$ is a geographic demand weight based on the sub-satellite latitude $\phi_e$ of the link midpoint, following population-weighted and GDP-weighted demand models used in LEO constellation design \cite{bhattacherjee2019network}. 

\subsubsection{Queuing}
Each ISL behaves as an M/M/1/K queue \cite{marcano2021queuing} with finite buffer capacity of $K$ packets and service rate $\mu_s = C_e / (B \cdot 8)$ packets/s, where $C_e$ is the link capacity and $B$ is the packet size. The link utilization is $\rho_e = (\lambda_e^{\text{fwd}} + \lambda_e^{\text{rev}}) / C_e$, the expected queue depth is $\mathbb{E}[N] = \rho/(1-\rho)$ for $\rho < 1$, and the packet loss probability under the finite-buffer model is defined as:
\begin{equation}
    P_{\text{loss}} = \frac{(1-\rho)\,\rho^{K}}{1 - \rho^{K+1}}.
    \label{eq:packet_loss}
\end{equation}

The end-to-end latency on link $e$ combines propagation latency ($\tau_e$), queuing latency ($\tau_{q,e}$), and per-hop processing latency ($\tau_\text{proc}$) as follows:
\begin{equation}
    \Lambda_e = \tau_e + \tau_{q,e} + \tau_{\text{proc}},
    \label{eq:e2e_latency}
\end{equation}

\subsection{Threat Model}
\label{subsec:threat_model}

We consider a bounded adversary that can (i) deploy one or more \emph{rogue} satellites in nearby orbits that impersonate legitimate constellation members \cite{kohli2026emma}, and (ii) compromise individual satellites through firmware exploitation or supply-chain insertion \cite{willbold2023space}, gaining control of the onboard routing and traffic forwarding functions. We also assume the adversary \emph{cannot} compromise the GEO-based ephemeris broadcast, which serves as a trusted anchor for position verification. Additionally, the security of \textit{feeder links} and \textit{user links} is beyond the scope of this paper.

The adversary aims to achieve one or more of the following: (i) impersonate a legitimate satellite, (ii) degrade network performance by dropping, delaying, or flooding packets, (iii) subvert routing by advertising false link costs or fabricating identities to attract or divert traffic, or (iv) exploit multi-operator peering to hijack cross-constellation traffic. We consider ten attack types \cite{salim2024cybersecurity, giuliari2021icarus} that achieve one or more of the above goals:
\begin{enumerate}
    \item \textbf{Rogue Satellite:} The adversary places an unauthorized transmitter that attempts to impersonate a legitimate satellite to spoof another legitimate satellite in the constellation.
    \item \textbf{Blackhole:} A compromised node drops all forward packets.
    \item \textbf{Grayhole:} A compromised node that drops packets selectively, giving the impression of congestion.
    \item \textbf{Sinkhole:} A compromised node that advertises false, low-cost routing to attract more traffic.
    \item \textbf{Wormhole:} Two colluding nodes replay packets via an out-of-band tunnel, reducing apparent end-to-end latency. 
    \item \textbf{Sybil:} A compromised node asserts multiple fake identities in the routing protocol, inflating the neighbor counts and routing table sizes. 
    \item \textbf{DoS Flooding:} A compromised node floods targeted ISLs with junk traffic, causing an overflow.
    \item \textbf{Rogue Sinkhole:} A composite attack wherein a rogue satellite launches a Sinkhole attack.
    \item \textbf{Rogue Sybil:} A composite attack wherein a rogue satellite launches a Sybil attack.
    \item \textbf{Inter-constellation Hijack:} A compromised boundary satellite advertises false cheap routes into the other operators' address space, attracting inbound cross-operator traffic. While conceptually similar to Sinkhole, the inter-constellation hijack attack is used specifically for cross-ISLs while other network attacks operate within the scope of a single constellation.
\end{enumerate}

\section{Proposed Detection Framework}
\label{sec:framework}

This section formulates the detection problem, defines the cross-layer feature space, presents three unsupervised detectors and their hyperparameters, describes the per-satellite training paradigm, and presents the detector-agnostic detection algorithm.

\subsection{Problem Formulation}
\label{subsec:problem}

Given the constellation graph $\mathcal{G}_t = (\mathcal{V}, \mathcal{E}_t)$ and threat model defined in Section~\ref{sec:network_threat}, for each active ISL $e \in \mathcal{E}_t$ let $\vect{x}_e^t \in \mathbb{R}^p$ denote a $p$-dimensional cross-layer feature vector that fuses physical-layer mismatch measurements with network-layer data. The detection objective is as follows: given a training set of normal feature vectors $\{\vect{x}_e^t\}_{t \in \mathcal{T}_{\text{train}}}$, learn a scoring function $\mathcal{A}(\vect{x}) : \mathbb{R}^p \to \mathbb{R}$ and a per-satellite threshold $\theta$ such that $\mathcal{A}(\vect{x}_e^t) > \theta$ flags link $e$ at time $t$ as anomalous. We enforce that the models must be (i)~\emph{unsupervised}, and (ii)~\emph{per-satellite}.

\subsection{Cross-Layer Feature Space}
\label{subsec:features}

Each ISL observation at timestep $t$ produces a $p=12$-dimensional feature vector $\vect{x}_e^t \in \mathbb{R}^{12}$ comprising three physical-layer mismatch features and nine network-layer features. Table~\ref{tab:features} enumerates the stated features.

\begin{table}[t]
\centering
\caption{Cross-layer feature set.}
\label{tab:features}
\renewcommand{\arraystretch}{1.2}
\resizebox{0.9\columnwidth}{!}{
\begin{tabular}{clcl}
\toprule
\textbf{\#} & \textbf{Feature} & \textbf{Symbol} & \textbf{Layer} \\
\midrule
1 & Range mismatch & $\delta_{\text{range}}$ & Physical \\
2 & Velocity mismatch & $\delta_{\text{vel}}$ & Physical \\
3 & Direction mismatch & $\delta_{\text{dir}}$ & Physical \\
\midrule
4 & Total throughput & $\Gamma_e$ & Network \\
5 & Link utilization & $\rho_e$ & Network \\
6 & Queue depth (sender) & $Q_{s_1}$ & Network \\
7 & Queue depth (receiver) & $Q_{s_2}$ & Network \\
8 & Queuing delay & $\tau_q$ & Network \\
9 & End-to-end latency & $\Lambda_e$ & Network \\
10 & Packet loss rate & $P_{\text{loss}}$ & Network \\
11 & Neighbor count & $n_e$ & Network \\
12 & Traffic asymmetry & $\eta_e$ & Network \\
\bottomrule
\end{tabular}}
\end{table}

\subsubsection{Physical-layer mismatch features}
The observing satellite $s_1$ compares its direct ISL measurements against ephemeris-predicted values derived from the GEO-broadcast state vectors of the claimed peer $s_2$. Three mismatch features are defined as follows:
\begin{equation}
    \delta_{\text{range}} = \frac{d_{\text{meas}} - d_{\text{eph}}}{d_{\text{eph}}}, \label{eq:range_mismatch}
\end{equation}
\begin{equation}
    \delta_{\text{vel}} = \frac{v_{r,\text{meas}} - v_{r,\text{eph}}}{|v_{r,\text{eph}}|},
    \label{eq:vel_mismatch}
\end{equation}
\begin{equation}
    \delta_{\text{dir}} = \arccos\!\left(\hat{\boldsymbol{\ell}}_{\text{meas}} \cdot \hat{\boldsymbol{\ell}}_{\text{eph}}\right),
    \label{eq:dir_mismatch}
\end{equation}
where $d_{\text{eph}} = \norm{\vect{r}_2^{\text{eph}} - \vect{r}_1}$ and $v_{r,\text{eph}} = (\vect{v}_2^{\text{eph}} - \vect{v}_1) \cdot \hat{\boldsymbol{\ell}}_{\text{eph}}$ are computed from ephemeris state vectors using Eq. \eqref{eq:range}-\eqref{eq:radvel}. Under normal conditions, these features are nearly zero, within the bounds of measurement noise. Under a rogue satellite attack, the measured quantities reflect the geometry to the rogue's true position while the ephemeris reflects the legitimate satellite's position, producing nonzero mismatch values that indicate a spoofing attack.

\subsubsection{Network-layer features}
We define nine features to capture traffic, queuing, and routing state. These include, total throughput $\Gamma_e = \Gamma_e^{\text{fwd}} + \Gamma_e^{\text{rev}}$ which is the sum of forward and reverse traffic, link utilization $\rho_e = \Gamma_e / C_e$ which is the ratio of total throughput to link capacity, traffic asymmetry $\eta_e = |\Gamma_e^{\text{fwd}} - \Gamma_e^{\text{rev}}| / \Gamma_e$ which captures directional imbalance, queue depths $Q_{s_1}, Q_{s_2}$, queuing delay $\tau_q$, end-to-end latency $\Lambda_e$ from Eq. \eqref{eq:e2e_latency}, packet loss rate $P_{\text{loss}}$ from Eq. \eqref{eq:packet_loss} which reflect congestion state, and neighbor count $n_e$ of the remote endpoint.

\subsection{Unsupervised Per-Satellite Detectors}
\label{subsec:detectors}

Each satellite $s_i \in \mathcal{V}$ hosts its own local detector model. This per-satellite design is motivated by spatial heterogeneity in the constellation since satellites at different orbital positions experience different neighbor counts, traffic volumes, and link geometries. The alternative of a globally trained model is not considered in this work since one individual model would need to accommodate the full range of normal variation across all satellites and their neighborhood, potentially reducing sensitivity to localized anomalies.

Model training is performed on the ground infrastructure by the constellation operator using normal ISL features collected from a constellation simulation and real data. Such ground-side simulation-based provisioning ensures that no model is trained on potentially compromised data from the orbit, since an onboard training approach would be vulnerable to an adversary that is already active during the data collection window. Each satellite is deployed with an initial detector, which may further be updated over-the-air if the constellation's topology and network behavior has changed over the course of its operation.

We evaluate three unsupervised anomaly detectors that operate on the same 12-dimensional feature space defined in the previous subsection. All detectors are trained exclusively per satellite and on normal data. The detection threshold of each satellite's model is set at the $(1 - \text{FPR}_{\text{target}})$ percentile of anomaly scores on a clean validation set for the corresponding satellite, with $\text{FPR}_{\text{target}} = 0.5\%$. Thus, no attacked samples are used to train the detectors.

\subsubsection{Cross-Layer Mahalanobis (CLM)}
The CLM detector computes the Mahalanobis distance of each observation from the normal feature centroid. During training on the normal set $X_{\text{train}} \in \mathbb{R}^{n \times p}$, the detector computes the feature-wise mean $\boldsymbol{\hat{\mu}}$ and standard deviation $\boldsymbol{\sigma}$, standardizes the data as $\tilde{X} = (X - \boldsymbol{\hat{\mu}}) / \boldsymbol{\sigma}$, and estimates the covariance matrix $\boldsymbol{\Sigma} = \text{Cov}(\tilde{X}) + 10^{-4}\mathbf{I}_p$, where the constant term $10^{-4}\mathbf{I}_p$ ensures invertibility.

The anomaly score for a test observation $\vect{x}$ is:
\begin{equation}
    D_M(\vect{x}) = \sqrt{(\tilde{\vect{x}} - \boldsymbol{\hat{\mu}})^\top \boldsymbol{\Sigma}^{-1} (\tilde{\vect{x}} - \boldsymbol{\hat{\mu}})},
    \label{eq:mahalanobis}
\end{equation}
where $\tilde{\vect{x}} = (\vect{x} - \boldsymbol{\hat{\mu}}) / \boldsymbol{\sigma}$. The detector flags $\vect{x}$ as anomalous when $D_M(\vect{x}) > \theta$.

\subsubsection{Isolation Forest (IF)}
Isolation Forest (IF) is an ensemble method that exploits the geometric property that anomalous points are both rare and structurally distinct, making them easier to separate from the bulk of the data. An ensemble of 200 binary isolation trees is constructed, each built by recursively selecting a random feature $j \in \{1, \ldots, p\}$ and a random split value $s \in [\min(x_j), \max(x_j)]$, partitioning the data until every sample occupies its own leaf node. The path length $h(\vect{x})$, which is the number of splits required to isolate observation $\vect{x}$, serves as the basis for anomaly scoring. Normal observations lie in dense regions of the feature space and require many splits to isolate, whereas anomalous observations that deviate from the nominal cluster are isolated in fewer splits. The anomaly score is defined as $s(\vect{x}) = 2^{-\mathbb{E}[h(\vect{x})]/c(n)}$, where $\mathbb{E}[h(\vect{x})]$ is the average path length across all trees and $c(n)$ is a normalization factor derived from the average path length of unsuccessful searches in a binary search tree of $n$ samples. Features are standardized prior to training and the detection threshold is calibrated on the validation set as done for CLM.

\subsubsection{Feedforward Autoencoder (AE)}
The Autoencoder (AE) used for this work is a symmetric feedforward network with encoder layers $12 \to 64 \to 32 \to 16$ and decoder layers $16 \to 32 \to 64 \to 12$. All layers are activated using Rectified Linear Unit activations. The model is trained to reconstruct normal feature vectors by minimizing Mean Squared Error (MSE) using the Adam optimizer at a learning rate of $10^{-3}$, batch size 256, and 50 epochs. At test time, the per-sample reconstruction error serves as the anomaly score. Anomalous observations (which deviate from the normal manifold learned during training) incur higher anomaly scores than normal observations that follow the training distribution.

\subsection{Detection Algorithm}
\label{subsec:algorithm}

Algorithm~\ref{algo:detection} describes the onboard detection procedure executed by each satellite. The algorithm is detector-agnostic.

\begin{algorithm}[t]
\caption{Onboard per-satellite cross-layer ISL anomaly detection.}
\label{algo:detection}
\DontPrintSemicolon
\KwIn{Pre-provisioned model $\mathcal{A}_i$, threshold $\theta_i$, active ISL set $\mathcal{N}_i^t$}
\KwOut{Detection decisions $\{y_e^t\}_{e \in \mathcal{N}_i^t}$, alert set $\mathcal{F}_i^t$}
\BlankLine
$\mathcal{F}_i^t \gets \emptyset$ \tcp*{Initialize alert set}
\ForEach{active ISL $e \in \mathcal{N}_i^t$}{
    \tcp{Step 1: Acquire raw measurements}
    $(d_{\text{meas}},\, v_{r,\text{meas}},\, \hat{\boldsymbol{\ell}}_{\text{meas}}) \gets \text{TerminalRead}(e)$ \;
    $(\Gamma_e,\, \rho_e,\, Q_{s_1},\, Q_{s_2},\, \tau_q,\, \Lambda_e,\, P_{\text{loss}},\, n_e,\, \eta_e) \gets \text{QueueState}(e)$ \;
    \BlankLine
    \tcp{Step 2: Compute physical-layer mismatch features}
    $(\vect{r}_2^{\text{eph}}, \vect{v}_2^{\text{eph}}) \gets \text{EphemerisLookup}(s_2(e),\, t)$ \;
    $\delta_{\text{range}} \gets (d_{\text{meas}} - d_{\text{eph}}) / d_{\text{eph}}$ \;
    $\delta_{\text{vel}} \gets (v_{r,\text{meas}} - v_{r,\text{eph}}) / |v_{r,\text{eph}}|$ \;
    $\delta_{\text{dir}} \gets \arccos(\hat{\boldsymbol{\ell}}_{\text{meas}} \cdot \hat{\boldsymbol{\ell}}_{\text{eph}})$ \;
    \BlankLine
    \tcp{Step 3: Assemble cross-layer feature vector}
    $\vect{x}_e^t \gets [\delta_{\text{range}}, \delta_{\text{vel}}, \delta_{\text{dir}}, \Gamma_e, \rho_e, Q_{s_1}, Q_{s_2}, \tau_q, \Lambda_e, P_{\text{loss}}, n_e, \eta_e]^\top$ \;
    \BlankLine
    \tcp{Step 4: Score and threshold}
    $a \gets \mathcal{A}_i(\vect{x}_e^t)$ \tcp*{Anomaly score}
    $y_e^t \gets \mathbf{1}[a > \theta_i]$ \tcp*{Binary decision}
    \If{$y_e^t = 1$}{
        $\mathcal{F}_i^t \gets \mathcal{F}_i^t \cup \{(e,\, a,\, t)\}$ \tcp*{Flag for operator}
    }
}
\Return $\{y_e^t\}_{e \in \mathcal{N}_i^t},\; \mathcal{F}_i^t$
\end{algorithm}

At each evaluation timestep, satellite $s_i$ iterates over its active ISL set $\mathcal{N}_i^t$ and performs the following four steps per link: First, acquire raw physical measurements from the ISL terminal and network-layer features from the local queue state. Second, compute the three physical-layer mismatch features by comparing measurements against ephemeris-predicted values. Third, assemble the 12-dimensional cross-layer feature vector $\vect{x}_e^t$. Fourth, score the observation using the pre-provisioned model $\mathcal{A}_i$ and compare against the threshold $\theta_i$. Links exceeding the threshold are flagged in the alert set $\mathcal{F}_i^t$, which is reported to the ground operator for further action. The scoring function $\mathcal{A}$ is instantiated based on the selected detector (CLM, IF, or AE), yielding a scalar anomaly score that increases with deviation from nominal behavior.

\section{Simulation and Dataset}
\label{sec:simulation}

This section describes the simulation parameters, dataset composition, and evaluation metrics.

\subsection{Simulation Parameters}
\label{subsec:sim_env}

\begin{figure}[t] 
    \centering
  \subfloat[Starlink]{%
       \includegraphics[width=\linewidth]{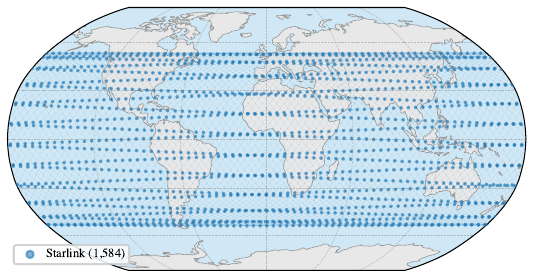}}\\
  \subfloat[Kuiper]{%
        \includegraphics[width=\linewidth]{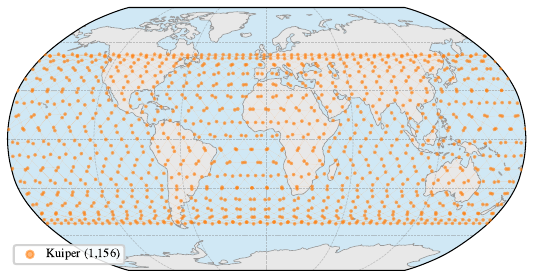}}\\
  \subfloat[Multi-operator]{%
        \includegraphics[width=\linewidth]{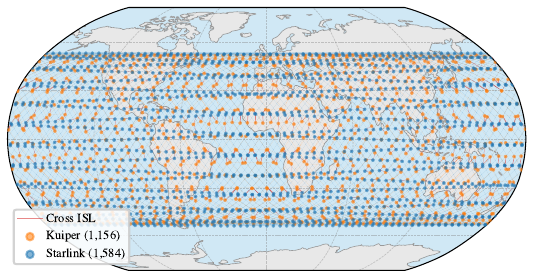}}
  \caption{Constellation snapshots at simulation initialization.}
  \label{fig:snapshots} 
\end{figure}

We construct a physics-grounded ISL simulation that generates realistic cross-layer features for three constellation scenarios: Starlink (1,584 satellites), Kuiper (1,156 satellites), and a theoretical Starlink-Kuiper multi-operator network (2,740 satellites). Satellite orbits are propagated using the SGP4 propagator in the ECEF frame \cite{rockwood2023generating}, with simulation epoch 2026-01-01 00:00~UTC spanning 24 hours at a sampling rate of 60~s. Figure \ref{fig:snapshots} depicts a snapshot of the constellation at the time of initialization.

Table~\ref{tab:constellation} lists the orbital and ISL parameters for each constellation scenario. The Starlink and Kuiper configurations correspond to their respective first-shell Federal Communications Commission (FCC) filings \cite{starlink_fcc, kuiper_fcc}. The multi-operator scenario combines both constellations with cross-constellation ISLs operating at reduced capacity to highlight the operators' preference for intra-satellite links. 

\begin{table}[t]
\centering
\caption{Constellation parameters per scenario.}
\label{tab:constellation}
\renewcommand{\arraystretch}{1.2}
\resizebox{\columnwidth}{!}{
\begin{tabular}{lccc}
\toprule
\textbf{Parameter} & \textbf{Starlink} & \textbf{Kuiper} & \textbf{Multi-op.} \\
\midrule
Altitude (km) & 550 & 630 & Both \\
Inclination (deg) & 53.0 & 51.9 & Both \\
Eccentricity & 0.0001 & 0.0001 & Both \\
Planes $\times$ Sats/plane & $72 \times 22$ & $34 \times 34$ & $106 \times -$ \\
Total satellites & 1,584 & 1,156 & 2,740 \\
ISL terminals per sat & 4 & 4 & 4 \\
Max ISL range (km) & 2,500 & 2,500 & 2,500 \\
ISL capacity (Gbps) & 100 & 100 & 25 \\
Carrier frequency (THz) & 193.4 & 193.4 & 193.4 \\
\bottomrule
\end{tabular}}
\end{table}

\textbf{Measurement noise.} Realistic sensor imperfections are modeled as additive noise on each ISL observable. These include range noise $\sigma_d = 5$~m, Doppler noise with fractional component $10^{-7} \cdot |f_d|$ and a floor 0.5~Hz, radial velocity noise $\sigma_v = 1$~mm/s, and pointing noise $\sigma_\ell = 1$~$\mu$rad \cite{hemmati2020near, wayne2024connecting}.

\textbf{Network parameters.} The simulation follows the model of Section~\ref{subsec:network_model} with congestion weight $\alpha = 0.5$, BGP peering penalty $\beta_{ij} = 2$~ms for cross-constellation links, base traffic arrival rate $\bar{\lambda} = 540$~Mbps \cite{brashears2024achieving}, queue capacity $K = 100$~packets, packet size $B = 1,500$~bytes, intra-constellation link capacity $C_e = 100$~Gbps \cite{brashears2024achieving}, cross-constellation capacity 25~Gbps, and per-hop processing delay $\tau_{\text{proc}} = 0.01$~ms. The geographic demand weight $g(\phi)$ is a sum-of-Gaussians population density proxy based on \cite{bhattacherjee2019network}:
\begin{equation}
    g(\phi) = 0.3 + 0.9\exp\!\left(\frac{-(\phi - 35)^2}{2 \cdot 18^2}\right) + 0.5\exp\!\left(\frac{-(\phi + 30)^2}{2 \cdot 15^2}\right),
    \label{eq:geo_weight}
\end{equation}
where $\phi$ is the sub-satellite latitude of the ISL midpoint in degrees. The first Gaussian ($\phi_0 = 35^{\circ}$N, $\sigma = 18^{\circ}$, weight 0.9) captures the densely populated northern-hemisphere band spanning Asia, Europe, and North America. The second Gaussian ($\phi_0 = 30^{\circ}$S, $\sigma = 15^{\circ}$, weight 0.5) covers the secondary southern-hemisphere population centers in southeast Brazil, South Africa, and southeast Australia, with a narrower spread and lower amplitude reflecting the smaller landmass and population at these latitudes. The constant baseline of 0.3 ensures that ISLs over polar and equatorial-oceanic regions still carry traffic. The function is mean-normalized so that $g(\phi)$ only redistributes traffic across latitudes without altering the aggregate volume set by $\bar{\lambda}$. Under this model, ISLs overpopulate mid-latitude regions carry approximately 1.5-1.7$\times$ of the average traffic, while polar ISLs carry approximately 0.3$\times$ of the average. While we follow this geographic demand model for our experiments, the proposed method attained nearly the same performance on alternative demand distributions.

\begin{table}[t]
\centering
\caption{Dataset size at 60~s sampling rate over 24 hours.}
\label{tab:dataset}
\renewcommand{\arraystretch}{1.2}
\resizebox{\columnwidth}{!}{
\begin{tabular}{lrrr}
\toprule
\textbf{Split} & \textbf{Starlink} & \textbf{Kuiper} & \textbf{Multi-operator} \\
\midrule
Training        & 1,822,679  & 1,328,970  & 4,483,361 \\
Validation      & 911,345     & 664,480     & 2,241,681 \\
Test (clean)    & 1,822,693  & 1,328,968  & 4,483,373 \\
Test (attacked) & 1,822,693  & 1,328,968  & 4,483,373 \\
\midrule
\textbf{Total}            & \textbf{6,379,410} & \textbf{4,651,386} & \textbf{15,691,788} \\
\bottomrule
\end{tabular}}
\end{table}

\begin{figure}[t]
    \centering
    \includegraphics[width=0.9\linewidth]{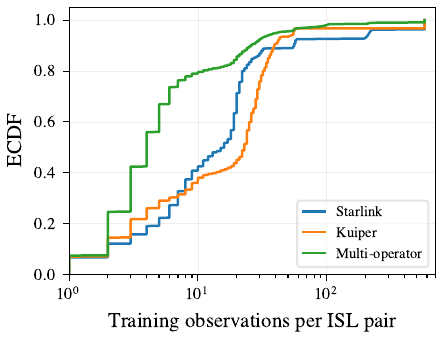}
    \caption{ECDF of number of observations per ISL.}
    \label{fig:isl_pair_observation}
\end{figure}

\begin{figure}[t]
    \centering
    \includegraphics[width=0.9\linewidth]{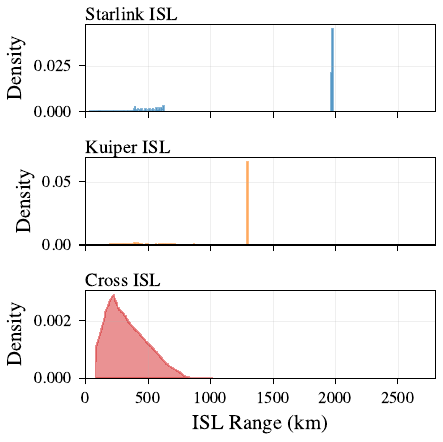}
    \caption{Range distribution by link type.}
    \label{fig:isl_range_distribution}
\end{figure}

\begin{figure}[t]
    \centering
    \includegraphics[width=\linewidth]{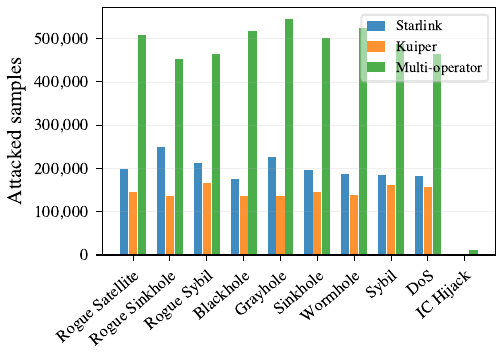}
    \caption{Number of attacked samples by type.}
    \label{fig:attack_distribution}
\end{figure}

\subsection{Attack Injection}
\label{subsec:attack_injection}

The ten attack types defined in Section~\ref{sec:network_threat} are injected into the test window of the adversarial dataset on a per-ISL basis. The injection procedure assigns each attacked ISL one of the ten attack types with equal probability, and one of three severity tiers in low (40\% of total instances), medium (40\% of total instances), or high (remaining 20\% of total instance), where severity refers to the extent of deviation from normal. Thus, low severity will be more difficult to detect than medium and high. Further, the inter-constellation hijack attack is restricted to cross-constellation ISLs in the multi-operator scenario. To avoid unrealistic constant-magnitude perturbations, the attack intensity oscillates sinusoidally with $\pm$30\% amplitude variation over the attack duration, ensuring that feature deviations vary in magnitude across time steps while still being persistently anomalous. To model a realistic cascading anomaly scenario, each attacked satellite's remaining ISLs have a 30\% probability of receiving the same attack type and severity tier. This produces correlated attack patterns across ISLs sharing a compromised endpoint, consistent with the adversary capabilities defined in the threat model.

\subsection{Dataset Construction}
\label{subsec:dataset}

Two parquet files are produced per scenario: \texttt{normal.parquet} contains the full simulation with no attacks, and \texttt{attacked.parquet} contains the adversarial test window with injected attacks.

The dataset is split temporally to test generalization of using partial simulation data to predict over the remainder of the simulation. We make the following splits:
\begin{itemize}
    \item \textbf{Training set}: Consists of the first 40\% of the normal samples from \texttt{normal.parquet}.
    \item \textbf{Validation set}: Consists of the next 20\% of the normal samples from \texttt{normal.parquet}.
    \item \textbf{Test set}: Consists of a normal subset comprising of the last 40\% of normal samples from \texttt{normal.parquet}, and an attack-injected copy of the same subset called \texttt{attacked.parquet}. 
\end{itemize}

Table~\ref{tab:dataset} reports the number of ISL observations per split. The multi-operator scenario produces the largest dataset due to the addition of cross-constellation ISLs. Derived features are computed after loading, ensuring that attack modifications to raw measurements propagate consistently to all derived quantities.

Fig.~\ref{fig:isl_pair_observation} shows the Empirical Cumulative Distribution Function (ECDF) of the number of observations per ISL in the three constellation scenarios while Fig.~\ref{fig:isl_range_distribution} depicts the ISL range distribution for inter-satellite and cross-ISL links observed in the simulation. Fig.~\ref{fig:attack_distribution} shows the attack distribution across the scenarios. Note that inter-constellation hijack constitutes only a small sample set on cross-ISLs (11,227 samples), since the other nine attack types have an equal probability on the cross-ISL connections in the test set.

\begin{table}[t]                    
\centering
\caption{Predictive performance of IF, AE and CLM on the simulation datasets.}
\label{tab:results}                            
\renewcommand{\arraystretch}{1.2}
\resizebox{\columnwidth}{!}{                                                                  \begin{tabular}{|c|c|c|c|c|c|c|c|}                                                            \hline
\textbf{Detector}       & \textbf{Scenario} & \textbf{P} & \textbf{R} & \textbf{F1} & \textbf{FPR} & \textbf{PR} & \textbf{ROC} \\   
\hline                                                                                                                               
\multirow{4}{*}{IF}  & Starlink          & 0.972 & 0.267 & 0.419 & 0.008 & 0.907 & 0.896 \\ \cline{2-8}
                   & Kuiper            & 0.974 & 0.265 & 0.416 & 0.007 & 0.899 & 0.884 \\ \cline{2-8}                              
                   & Multi             & 0.979 & 0.276 & 0.430 & 0.006 & 0.892 & 0.873 \\ \cline{2-8}                              
                   & \cellcolor{yellow!30}Average & \cellcolor{yellow!30}0.975 & \cellcolor{yellow!30}0.269 &                      
\cellcolor{yellow!30}0.422 & \cellcolor{yellow!30}0.007 & \cellcolor{yellow!30}0.899 & \cellcolor{yellow!30}0.884 \\ \hline          
\multirow{4}{*}{AE}  & Starlink          & 0.993 & 0.941 & 0.967 & 0.006 & 0.992 & 0.990 \\ \cline{2-8}                              
                   & Kuiper            & 0.993 & 0.943 & 0.967 & 0.007 & 0.990 & 0.987 \\ \cline{2-8}                              
                   & Multi             & 0.993 & 0.888 & 0.938 & 0.006 & 0.986 & 0.982 \\ \cline{2-8}                              
                   & \cellcolor{yellow!30}Average & \cellcolor{yellow!30}0.993 & \cellcolor{yellow!30}0.924 &                      
\cellcolor{yellow!30}0.957 & \cellcolor{yellow!30}\textbf{0.006} & \cellcolor{yellow!30}0.990 & \cellcolor{yellow!30}0.986 \\ \hline 
\multirow{4}{*}{CLM} & Starlink          & 0.994 & 0.996 & 0.995 & 0.006 & 1.000 & 1.000 \\ \cline{2-8}                              
                   & Kuiper            & 0.993 & 0.995 & 0.994 & 0.007 & 1.000 & 1.000 \\ \cline{2-8}                              
                   & Multi             & 0.994 & 0.948 & 0.970 & 0.006 & 0.998 & 0.997 \\ \cline{2-8}                              
                   & \cellcolor{yellow!30}Average & \cellcolor{yellow!30}\textbf{0.994} & \cellcolor{yellow!30}\textbf{0.979} &    
\cellcolor{yellow!30}\textbf{0.986} & \cellcolor{yellow!30}\textbf{0.006} & \cellcolor{yellow!30}\textbf{0.999} &                    
\cellcolor{yellow!30}\textbf{0.999} \\ \hline                   \end{tabular}}
\end{table} 

\begin{figure*}[t] 
    \centering
  \subfloat[Starlink]{%
       \includegraphics[width=0.33\linewidth]{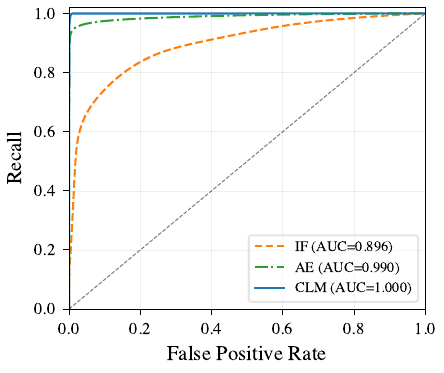}}
  \subfloat[Kuiper]{%
        \includegraphics[width=0.33\linewidth]{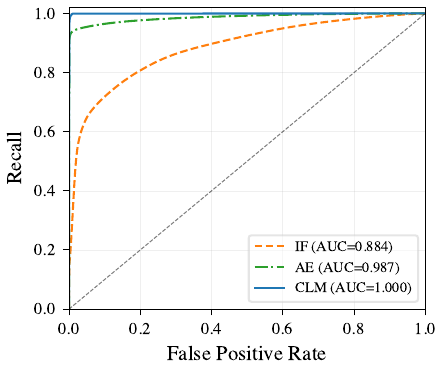}}
  \subfloat[Multi-operator]{%
        \includegraphics[width=0.33\linewidth]{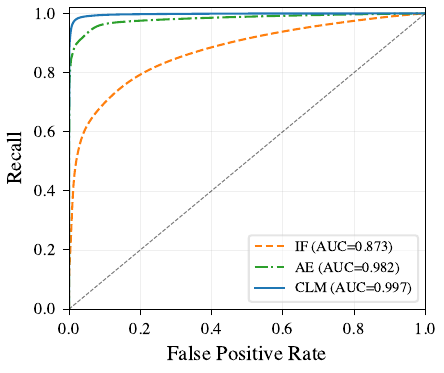}}
  \caption{ROC-AUC per scenario.}
  \label{fig:roc} 
\end{figure*}

\begin{figure*}[t] 
    \centering
  \subfloat[Starlink]{%
       \includegraphics[width=0.33\linewidth]{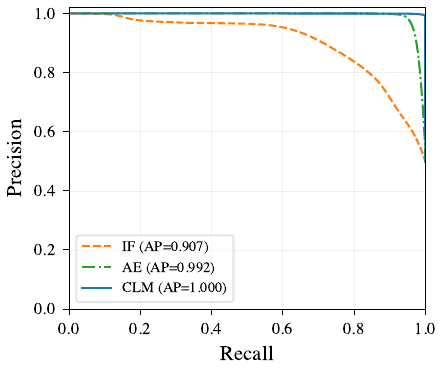}}
  \subfloat[Kuiper]{%
        \includegraphics[width=0.33\linewidth]{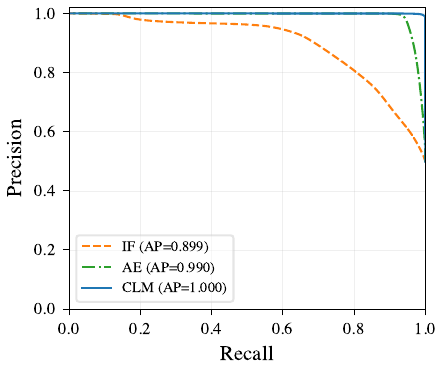}}
  \subfloat[Multi-operator]{%
        \includegraphics[width=0.33\linewidth]{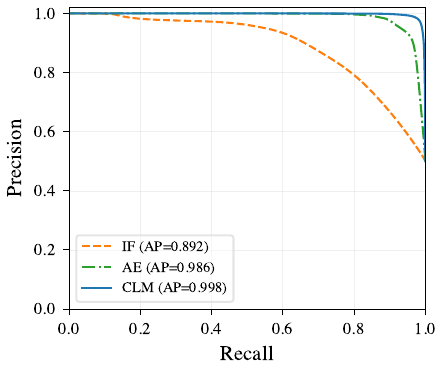}}
  \caption{PR-AUC per scenario.}
  \label{fig:pr} 
\end{figure*}

\subsection{Evaluation Metrics}
\label{subsec:metrics}

We report six standard binary classification metrics computed on the combined test set consisting of clean and attacked samples, including precision ($P$), recall ($R$), F1-score, FPR, area under the Receiver Operating Characteristic curve (ROC-AUC), and area under the Precision-Recall curve (PR-AUC). Among these, Recall and FPR are the primary metrics. Recall measures the fraction of attacks detected, while FPR measures the false alarm rate on clean traffic.

We additionally report per-attack recall and per-severity recall. For available prior-work baselines on physical-layer security, we compute \emph{scoped recall} restricted to the attack types each method was designed to detect, ensuring a fair comparison that does not penalize methods for attacks outside their design scope.

\begin{figure}[t] 
    \centering
  \subfloat[Starlink]{%
       \includegraphics[width=\linewidth]{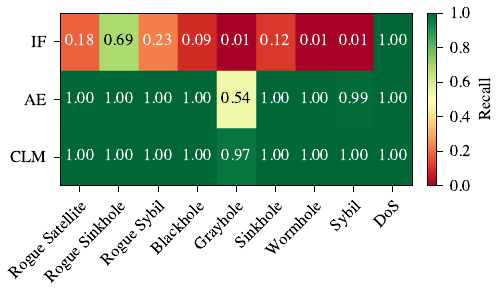}} \\
  \subfloat[Kuiper]{%
        \includegraphics[width=\linewidth]{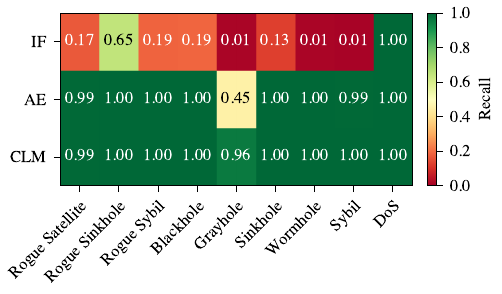}} \\
  \subfloat[Multi-operator]{%
        \includegraphics[width=\linewidth]{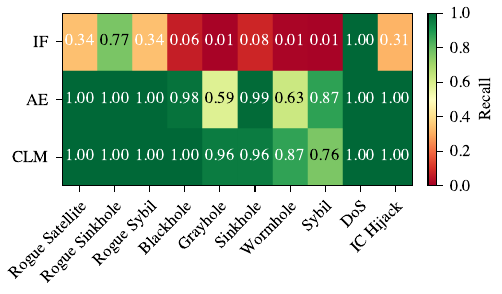}}
  \caption{Attack recall heatmap per scenario.}
  \label{fig:heatmap} 
\end{figure}

\begin{figure}[t] 
    \centering
  \subfloat[IF]{%
       \includegraphics[width=0.95\linewidth]{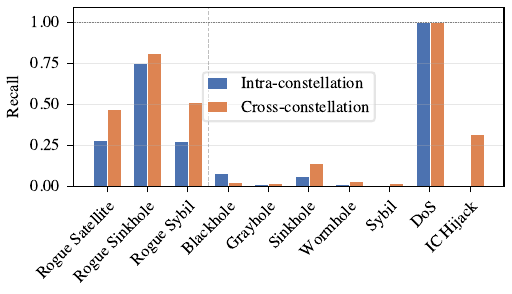}} \\
  \subfloat[AE]{%
        \includegraphics[width=0.95\linewidth]{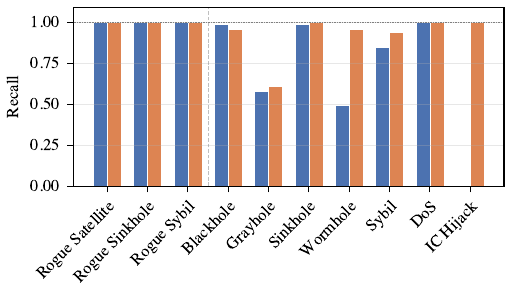}} \\
  \subfloat[CLM]{%
        \includegraphics[width=0.95\linewidth]{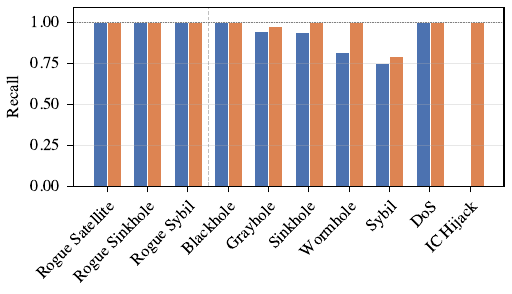}}
  \caption{Intra vs cross-constellation attack recall in the multi-operator constellation.}
  \label{fig:intravsmulti} 
\end{figure}

\begin{figure}[t] 
    \centering
  \subfloat[Starlink]{%
       \includegraphics[width=0.7\columnwidth]{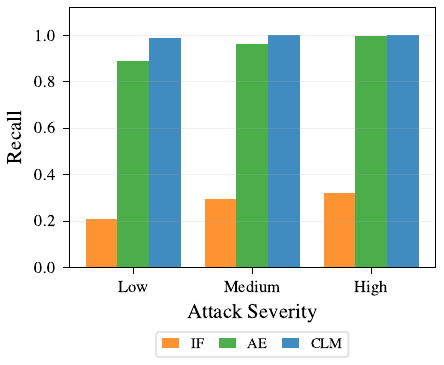}} \\
  \subfloat[Kuiper]{%
        \includegraphics[width=0.7\columnwidth]{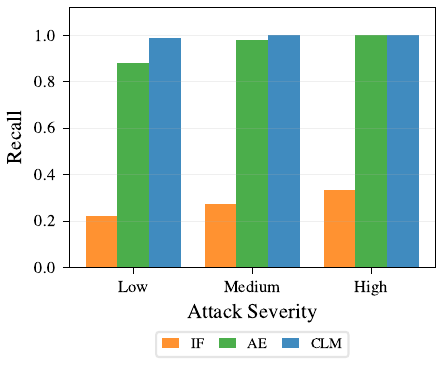}} \\
  \subfloat[Multi-operator]{%
        \includegraphics[width=0.7\columnwidth]{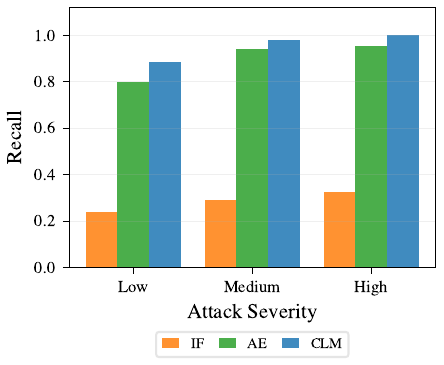}} \\
  \caption{Recall for different severity of attacks.}
  \label{fig:severity} 
\end{figure}

\section{Numerical Results}
\label{sec:results}

This section presents the performance of the three unsupervised detectors across all constellation scenarios, analyzes per-attack and per-severity recall, and compares against available prior-work baselines.

\subsection{Overall Detection Performance}

Table~\ref{tab:results} reports the aggregate metrics for IF, AE, and CLM across all three constellation scenarios. CLM achieves the highest average recall (97.9\%) and F1-score (98.6\%) while maintaining a false-positive rate of 0.6\%, matching AE's FPR. The slight increase in test FPR over $\text{FPR}_\text{target}=0.5\%$ tuned on the validation set is due to a distribution shift. AE achieves a recall of 92.4\% average but falls short of CLM by 5.5\% points, with the gap widening in the multi-operator scenario (88.8\% vs. 94.8\% for AE vs. CLM). IF exhibits high precision (97.5\%) but very low recall (26.9\%), indicating that it detects only the most extreme anomalies while missing the more subtle attacks. The ROC-AUC and PR-AUC curves in Fig.~\ref{fig:roc} and~\ref{fig:pr}, respectively, confirm these trends across the full threshold range. CLM achieves near-perfect ROC-AUC (0.997) and PR-AUC (0.998) on all scenarios. AE curves are close behind, while IF shows a clear separation, reflecting its inability to capture the full range of attack signatures at any single threshold.

All detectors show a modest performance degradation in the multi-operator scenario. CLM recall drops from 99.6\% (Starlink) and 99.5\% (Kuiper) to 94.8\% (multi-operator). This is attributable to the heterogeneity introduced by cross-ISLs, which operate at reduced capacity and exhibit different traffic patterns than intra-constellation ISLs, widening the normal feature distribution that each per-satellite model must accommodate.

\begin{figure*}
    \centering
    \includegraphics[width=0.95\linewidth]{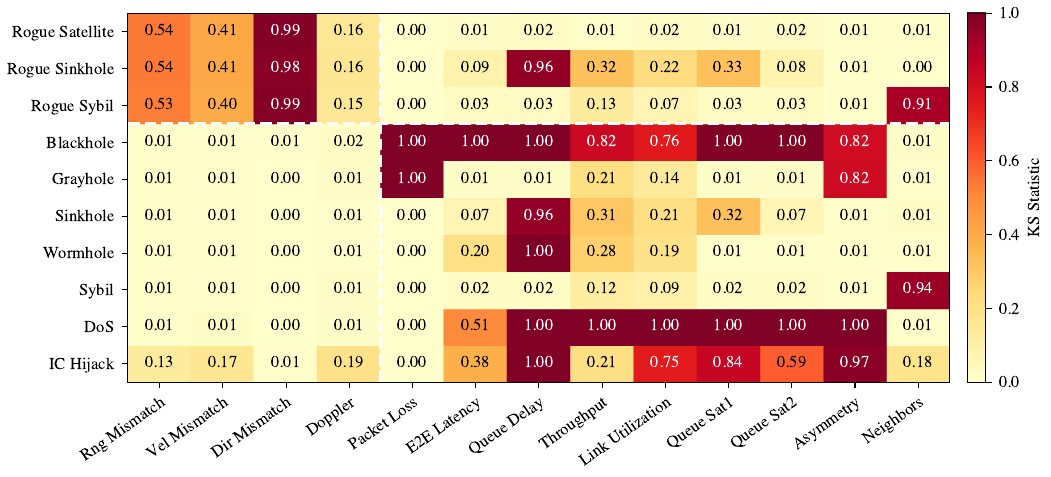}
    \caption{Attack-feature signatures.}
    \label{fig:feature_signature}
\end{figure*}

\subsection{Per-Attack Recall}

Fig.~\ref{fig:heatmap} shows the per-attack-type recall for each detector and scenario. CLM achieves perfect or near-perfect recall (99.9\%) on all threats except Grayhole in Starlink and Kuiper. The hardest attack for CLM is Grayhole (96.9\% on Starlink, 95.9\% on Kuiper), which is by design the most challenging to detect as it produces feature deviations that overlap with natural congestion.

In the multi-operator scenario, CLM's recall drops for Sybil (76.3\%) and Wormhole (86.8\%). Sybil attacks inflate only the neighbor count feature $n_e$, and the higher variability of $n_e$ on cross-constellation ISLs reduces the per-satellite model's ability to distinguish attack-induced inflation from normal behavior. Wormhole attacks reduce apparent latency, but cross-constellation ISLs already exhibit higher latency variance due to the more variable link distances, again reducing separability. IF's poor recall concentrates on network-only attacks and it achieves perfect recall only on DoS flooding, the most extreme attack in terms of feature magnitude. While AE does at par or worse than CLM in 8 attack types, it outperforms CLM in Sinkhole and Sybil attacks in this scenario. Further, Fig. \ref{fig:intravsmulti} highlights the attack recall comparison between intra and cross constellation samples within multi-operator. AE and CLM perform better in cross-constellation attack scenarios due to high link characteristic volatility. 

\subsection{Per-Severity Recall}

Fig.~\ref{fig:severity} shows recall stratified by severity tier. CLM maintains over 98.8\% recall on medium and high severity across Starlink and Kuiper, with low-severity recall at 98.8\% (Starlink) and 98.7\% (Kuiper). In the multi-operator scenario, CLM's low-severity recall drops to 88.6\%, while medium (98.1\%) and high (99.97\%) remain high. This confirms that the multi-operator performance gap is driven primarily by low-severity attacks on cross-ISLs. AE shows a similar pattern but with a larger degradation at low severity, particularly in the multi-operator scenario. IF shows uniformly low recall across all severity tiers except for the most extreme perturbations seen in high severity.

\subsection{Feature Signatures and Correlation}

Fig.~\ref{fig:feature_signature} visualizes the feature-level signatures of each attack type in the multi-operator scenario. Physical-layer attacks (rogue satellite, rogue sinkhole, rogue sybil) produce strong deviations in $\delta_{\text{range}}$, $\delta_{\text{vel}}$, and $\delta_{\text{dir}}$, while leaving network features at normal levels. Network-layer attacks such as Blackhole, Sinkhole, and DoS perturb throughput, utilization, and queue features while physical feature signatures remain near zero for these attacks. The composite rogue-class attacks activate both feature subsets simultaneously, confirming the benefit of cross-layer detection for LEO security.




\begin{figure}
    \centering
    \includegraphics[width=0.85\linewidth]{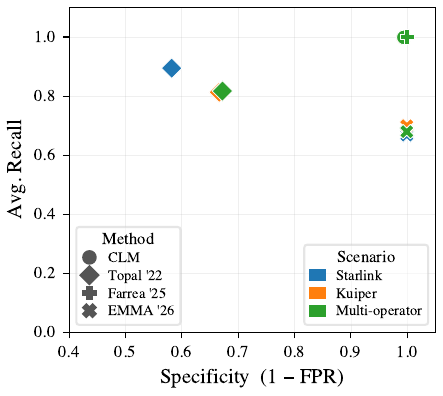}
    \caption{Prior work Recall vs FPR for prior works on physical-layer security of LEO ISLs and CLM on attack variants of Rogure Satellite.}
    \label{fig:prior_work}
\end{figure}

\subsection{Detector Complexity}

Onboard ISL security solutions must operate within strict energy budgets and must not introduce overheads that degrade the low-latency services ISLs are designed to provide. CLM requires only 169 stored parameters spanning a $12 \times 12$ inverse covariance matrix constituting 144 parameters, a mean vector of 12 parameters, a standard deviation vector of 12 parameters, and a scalar threshold taking 1 parameter. Further, it performs 337 floating-point operations per observation, dominated by a single matrix-vector product. This makes CLM the lightest detector by over two orders of magnitude fewer parameters than IF and over 40$\times$ fewer FLOPs than AE. CLM performs 1,348 FLOPs per observation, a negligible load for any onboard processor. IF stores approximately 23,500 node parameters and requires $\sim$1,550 comparisons per observation. While inference is fast, the storage footprint is 139$\times$ larger than CLM for no detection benefit. AE requires 6,876 parameters and $\sim$13,700 FLOPs per observation, which is 41$\times$ more compute than CLM per observation. CLM's combination of detection performance, low computational cost, and ease of tuning makes it the clear choice for large-scale onboard deployment that works within the limited energy budget of satellites.

\subsection{Comparison with Prior Work}

As discussed in Section \ref{sec:related_work}, the lack of physical, network, and cross-layer datasets for LEO constellations and ISLs has limited the amount of prior work on LEO constellation security. Among the related works covered in Table \ref{tab:related}, the physical-layer methods are, by design, applicable only to rogue-class attacks and none of the network-layer methods propose detectors. This subsection presents a comparison of the physical-layer approaches \cite{kohli2026emma,farrea2025zero,topal2022physical} against our best model (CLM), evaluated in our simulation. Fig.~\ref{fig:prior_work} presents this comparison with a recall vs sensitivity (1-FPR) scatter plot for the three constellation deployments. \cite{farrea2025zero} achieves 100\% recall at 0\% FPR on rogue-class attacks due to its cryptographic authentication approach. \cite{topal2022physical} achieves 89.5\% recall but at an FPR of 41.8\%, which would be operationally prohibitive in a constellation. \cite{kohli2026emma} achieves near-zero FPR, but at an average recall of around 70\%. In comparison to the three prior works, CLM achieves high recall (99.6\% on rogue-class attacks) at FPR $\leq$ 0.7\%, while also having the provision to detect network-layer and cross-layer attacks. This makes our proposed framework objectively better than the available research on LEO constellation security.

\section{Limitations and Future Works}
\label{sec:limitations}

While Section \ref{sec:results} demonstrate strong detection performance, several limitations should be noted. First, all results are derived from a physics-grounded simulation rather than an operational LEO deployment. Although the simulation incorporates SGP4 orbit propagation, realistic measurement noise, and congestion-aware traffic modeling, real ISL data may exhibit additional complexities such as atmospheric scintillation on near-horizon links, hardware aging effects, and transient faults that are not captured in the current model. Validation on real ISL data, when it becomes available, remains an important next step. Until then, our simulation dataset is the only available dataset on cross-layer simulation of realistic, dense and multi-operator constellations. Second, the threat model assumes a non-adaptive adversary whose attack parameters are fixed at injection time, with sinusoidal intensity variation as the only temporal dynamics. A sophisticated adversary that observes the detector's behavior and adapts its attack to remain below the detection threshold (\ie, an evasion attack) is not considered. Extending the framework to adversarial robustness, for example through min-max training or game-theoretic formulations, is a direction for future work. And third, the framework assumes that the GEO-based ephemeris broadcast is trusted. If the ephemeris itself is compromised, physical-layer mismatch features lose their reference baseline. A defense-in-depth approach that cross-validates ephemeris from multiple independent sources would strengthen this assumption.

\section{Conclusion}
This paper presented a cross-layer behavioral fingerprinting framework for detecting physical and network-layer attacks on ISLs in LEO mega-constellations and presented the first simulation dataset on dense and multi-operator constellation scenarios. The framework fused three physical-layer mismatch features derived from ephemeris comparison with nine network-layer features into a 12-dimensional per-ISL feature vector, enabling unsupervised, per-satellite anomaly detection that required no attack labels and no inter-satellite coordination. Performance was evaluated across Starlink (1,584 satellites), Kuiper (1,156 satellites), and a multi-operator scenario (2,740 satellites), injecting ten attack types spanning physical deception, network manipulation, cross-layer composites, and multi-operator exploitation. The proposed Mahalanobis detector achieved the strongest performance with 99.6\% recall on Starlink, 99.5\% on Kuiper, and 94.8\% on the multi-operator constellation, with FPR below 0.7\%. Comparison with prior physical-layer authentication methods confirmed that cross-layer fusion preserves rogue-detection capability while extending coverage to the full attack taxonomy.

The results establish two findings. First, cross-layer feature fusion enables the detection of physical, network and composite attacks. Second, a lightweight statistical detector such as Mahalanobis distance is sufficient for this task and outperforms both a tree-based ensemble and a deep AE, suggesting that the cross-layer feature space provides strong separability without requiring complex model architectures. Future work may validate our approach on real ISLs as operational data becomes available, extend to adaptive adversary models, and investigate ephemeris compromise.

\section*{Acknowledgement}
This paper was prepared with the assistance of an LLM in improving language clarity and editing. All technical content, research design, simulation implementation, experimental evaluation, and scientific conclusions are solely the work of the authors. The authors have reviewed and verified all AI-assisted content for correctness and accuracy, and take full responsibility for the content of this paper.

\bibliographystyle{IEEEtran}
\bibliography{references}

\end{document}